\documentclass[acmtog]{acmart}

\usepackage{booktabs} 

\usepackage{enumitem}
\usepackage{multirow} 
\usepackage{soul}
\sethlcolor{yellow}

\usepackage[ruled]{algorithm2e} 

\SetAlFnt{\small}
\SetAlCapFnt{\small}
\SetAlCapNameFnt{\small}
\SetAlCapHSkip{0pt}

\copyrightyear{2025}
\acmYear{2025}
\setcopyright{acmlicensed}\acmConference[SA Conference Papers '25]{SIGGRAPH Asia 2025 Conference Papers}{December 15--18, 2025}{Hong Kong, Hong Kong}
\acmBooktitle{SIGGRAPH Asia 2025 Conference Papers (SA Conference Papers '25), December 15--18, 2025, Hong Kong, Hong Kong}
\acmDOI{10.1145/3757377.3764009}
\acmISBN{979-8-4007-2137-3/2025/12}

\begin{document}
\title{AniMaker: Multi-Agent Animated Storytelling with MCTS-Driven Clip Generation}

\author{Haoyuan Shi}
\affiliation{
\institution{Harbin Institute of Technology, Shenzhen}
\country{China}}
\email{g1016015592@gmail.com}

\author{Yunxin Li}
\affiliation{
\institution{Harbin Institute of Technology, Shenzhen}
\country{China}}
\email{liyunxin987@163.com}

\author{Xinyu Chen}
\affiliation{
\institution{Harbin Institute of Technology, Shenzhen}
\country{China}}
\email{chenxinyuhitsz@163.com}

\author{Longyue Wang}
\affiliation{
\institution{Alibaba International Digital Commerce, Hangzhou}
\country{China}}
\email{vincentwang0229@gmail.com}

\author{Baotian Hu}
\authornote{Corresponding author: Baotian Hu.}
\affiliation{
\institution{Harbin Institute of Technology, Shenzhen}
\country{China}}
\email{hubaotian@hit.edu.cn}

\author{Min Zhang}
\affiliation{
\institution{Harbin Institute of Technology, Shenzhen}
\country{China}}
\email{zhangmin2021@hit.edu.cn}

\renewcommand\shortauthors{Haoyuan Shi, Yunxin Li, Xinyu Chen, et al}

\begin{abstract}
Despite rapid advancements in video generation models, generating coherent, long-form storytelling videos that span multiple scenes and characters remains challenging. Current methods often rigidly convert pre-generated keyframes into fixed-length clips, resulting in disjointed narratives and pacing issues. Furthermore, the inherent instability of video generation models means that even a single low-quality clip can significantly degrade the entire output animation's logical coherence and visual continuity. To overcome these obstacles, we introduce AniMaker, a multi-agent framework enabling efficient multi-candidate clip generation and storytelling-aware clip selection, thus creating globally consistent and story-coherent animation solely from text input. The framework is structured around specialized agents, including the Director Agent for storyboard generation, the Photography Agent for video clip generation, the Reviewer Agent for evaluation, and the Post-Production Agent for editing and voiceover, collectively realizing multi-character, multi-scene animation. Central to AniMaker's approach are two key technical components: MCTS-Gen in Photography Agent, an efficient Monte Carlo Tree Search (MCTS)-inspired strategy that intelligently navigates the candidate space to generate high-potential clips while optimizing resource usage; and AniEval in Reviewer Agent, the first framework specifically designed for multi-shot animation evaluation, which assesses critical aspects such as story-level consistency, action completion, and animation-specific features by considering each clip in the context of its preceding and succeeding clips. Experiments demonstrate that AniMaker achieves superior quality as measured by popular metrics including VBench and our proposed AniEval framework, while significantly improving the efficiency of multi-candidate generation, pushing AI-generated storytelling animation closer to production standards. Code and data for this paper are at https://animaker-dev.github.io/ 
\end{abstract}

%
%


%
%


\begin{teaserfigure}
    \centering
    \includegraphics[width=0.98\textwidth]{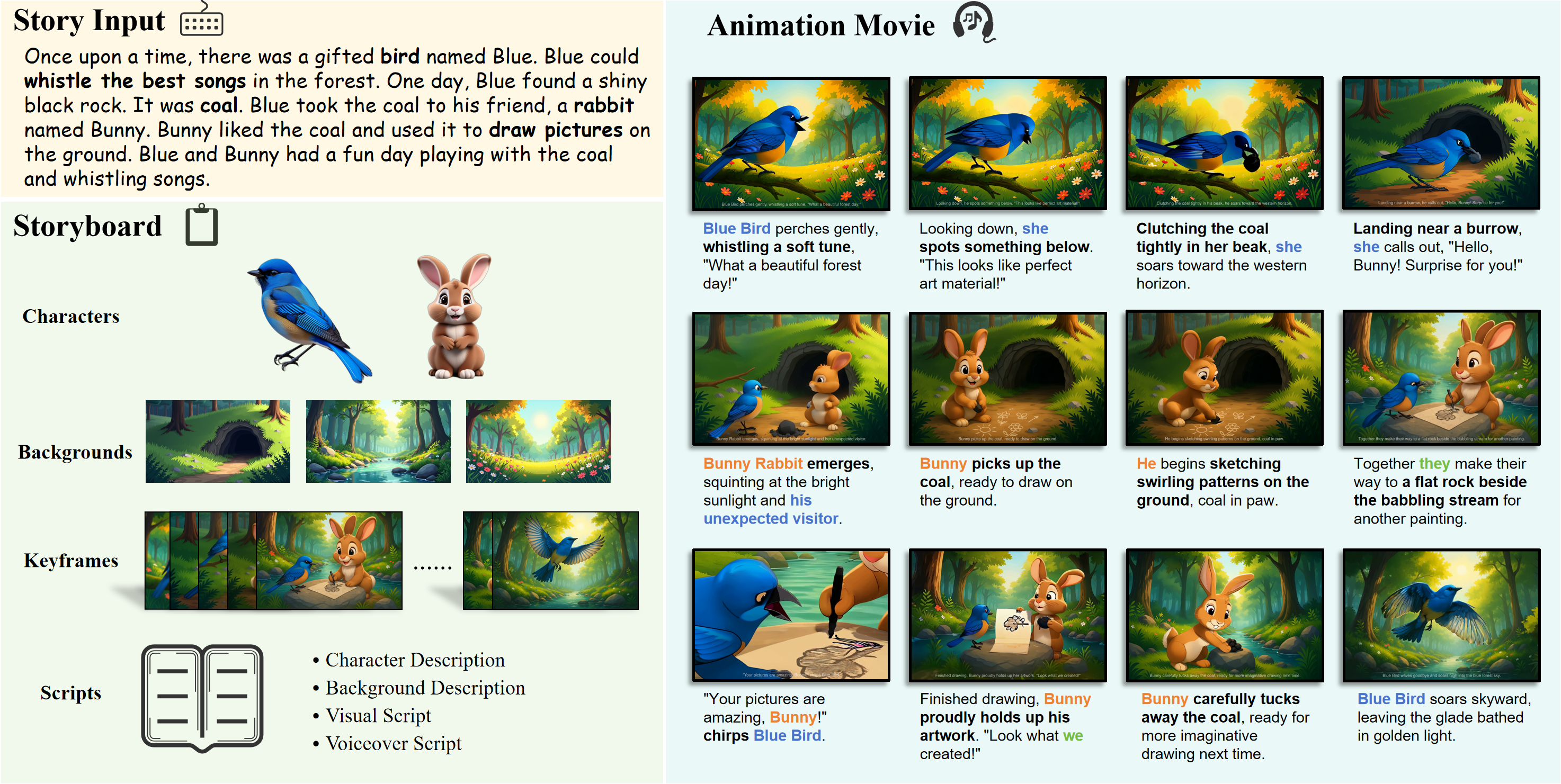}
    \caption{A visual example of AniMaker generating compelling storytelling animation from narrative text. Our framework maintains consistent character appearance across scenes while delivering high-quality action representation for complex sequences. AniMaker seamlessly integrates adaptive shot scheduling with smooth transitions, ensuring narrative coherence throughout the animation.
}
    \label{fig:intro-case}
\end{teaserfigure}

\maketitle

\captionsetup[figure]{skip=5pt}
\captionsetup[table]{skip=5pt}
\setlength{\textfloatsep}{8pt} 
\setlength{\floatsep}{8pt} 
\setlength{\intextsep}{8pt} 

\vspace{-0.5em}
\section{Introduction}

The emergence of large-scale video generation models \cite{yang2024cogvideox, kong2024hunyuanvideo, liu2024sora} has ignited significant interest in storytelling videos. In response, a new wave of methods \cite{kim2024fifo, wu2024mind, dalal2025one, guo2025long, xu2025filmagent} including MINT Video and TTT-Video has emerged, aiming to generate the entire video in a single pass. However, these methods still face significant challenges when generating long videos spanning multiple scenes and characters, particularly in maintaining visual continuity, narrative coherence, and non-repetitive content. In contrast, recent advancements in single-clip video generation, exemplified by models like Vidu\cite{bao2024vidu}, Wan \cite{wan2025wan}, and Pika \cite{PikaLabsWebsite}, have pushed generated video clips closer to cinematic quality. Due to shorter clip duration, these models utilize textual and visual prompts directly and efficiently, improving motion quality, semantic fidelity, and consistency with reference images. Consequently, utilizing these models to generate a storytelling video with multiple clips appears highly promising.

Existing storytelling video frameworks \cite{lin2023videodirectorgpt, he2024dreamstory, xie2024dreamfactory, li2024anim, xu2025mm, wu2025automated}, which implement the composition of multiple video clips, generally follow a fixed pipeline: script → keyframes → video clips → final video composition. While this modular format can effectively generate multi-shot long videos, it also introduces several limitations: Firstly, existing methods typically map keyframes to fixed-length clips. It forms a rigid and fragmented video construction, leading to disjointed transitions and unnatural pacing, severely hindering the expressive continuity crucial for complex or extended actions \cite{jiang2021camera}. Secondly, due to the inherent instability of video generation models, a single flawed clip can noticeably degrade the overall video quality.

To mitigate fragmentation, an intuitive approach is generating continuous clips conditioned on prior frames, but this compounds error propagation and quality degradation. Drawing parallels with professional filmmaking, we identify that existing methods overlook \textbf{Best-of-N Sampling}—generating multiple clip candidates and selecting the best ones. However, implementing this faces two challenges: prohibitive computational costs and inadequate automated evaluation mechanisms. Generating and evaluating multiple candidate clips per shot is computationally intensive, often relying on expensive commercial APIs or prolonged GPU inference. Current evaluation metrics like VBench \cite{xing2024survey} only assess individual clips and their internal consistency, neglecting critical elements such as cross-clip coherence, sequential motion quality, and animation-specific qualities in storytelling animation. 

To address these challenges, we introduce \textbf{AniMaker}, a multi-agent framework with MCTS-driven clip generation. This framework mirrors professional production \cite{xu2025filmagent}, including the \textbf{Director Agent} for storyboards construction, \textbf{Photography Agent} for video clip generation, \textbf{Reviewer Agent} for evaluating the quality of video clip candidates, and \textbf{Post-Production Agent} for editing and assembling the entire sequence of videos. These agents collaborate to enable automated multi-character, multi-scene storytelling without manual pre- or post-processing. The core Photography and Reviewer agents interact under a MCTS-Gen scheme during clip generation. Specifically, based on Monte Carlo Tree Search (MCTS), \textbf{MCTS-Gen} offers an efficient strategy for navigating the vast candidate space of video generation. It strikes a balance between broad exploration and computational efficiency by intelligently allocating more generation opportunities to promising clips while encouraging the exploration of unexplored regions. For Reviewer Agent, we introduce a comprehensive evaluation framework \textbf{AniEval}, specifically designed for multi-shot storytelling \cite{zhuang2025vistorybench} animation. AniEval advances beyond metrics like VBench by implementing retrospective evaluation with cross-clip contextual references. It evaluates critical dimensions—story consistency, action completion, and animation-specific attributes—by analyzing each clip in the context of its preceding and succeeding clips. This contextual assessment serves as the quality evaluation mechanism for video-clip nodes within the MCTS-Gen framework.

We conduct extensive experiments on the dataset constructed from TinyStories \cite{eldan2023tinystories}, featuring complex interaction with multiple characters across diverse backgrounds.
Experimental results demonstrate AniMaker's superior performance across VBench, AniEval, and human evaluation, with significantly improved multi-candidate generation efficiency. This validates AniMaker's effectiveness in bringing AI-generated storytelling videos closer to production-grade quality (as shown in Figure~\ref{fig:intro-case}). With our current work successfully targeting simple storytelling, we believe further development is essential to produce more complex and sophisticated animation styles.

In summary, our main contributions are:
\begin{itemize}
    \item We propose AniMaker, a fully automated multi-agent framework for generating coherent, multi-character, multi-scene animation from textual stories. The framework features MCTS-Gen, a novel search-based generation strategy that efficiently balances exploration and resource usage.
    \item We develop AniEval, the first comprehensive evaluation framework specifically designed for multi-shot storytelling animation. It provides context-aware assessment by analyzing clips in relation to their surrounding content.
    \item We experimentally demonstrate Animaker's superior performance across all evaluation frameworks—achieving best scores across all keyframe evaluation metrics, ranking 1st in VBench, 14.6\% higher scores in our AniEval, and better human ratings (3.22 versus 2.07).
\end{itemize}

\begin{figure*}[t]
    \vspace{-0.5em}
    \centering
    \includegraphics[width=0.99\textwidth]{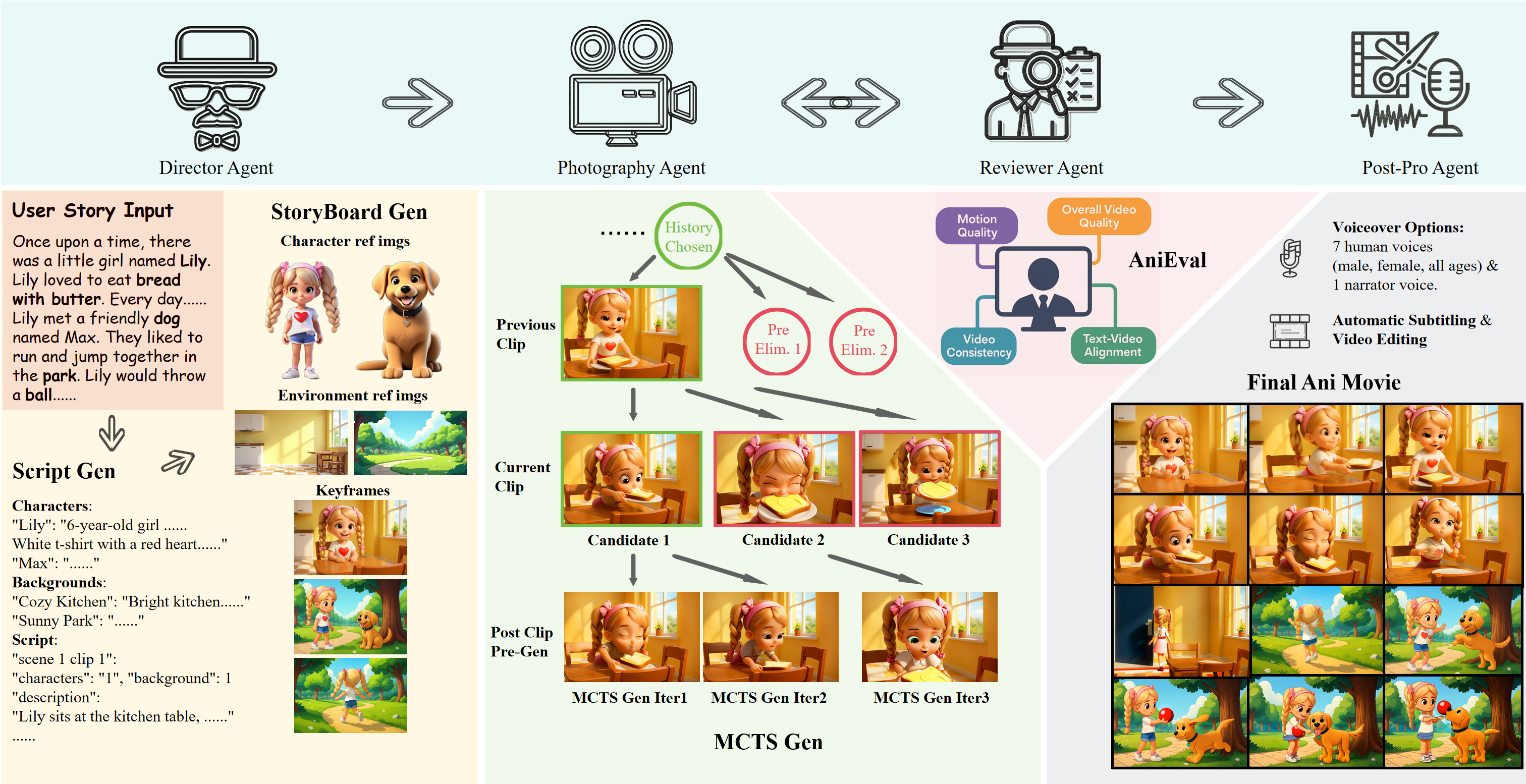}
    \caption{The overall architecture of our AniMaker framework. Given a story input, Director Agent creates detailed scripts and storyboards with reference images. Photography Agent generates candidate video clips using MCTS-Gen, which optimizes exploration-exploitation balance. Reviewer Agent evaluates clips with our AniEval assessment system. Post-production Agent assembles selected clips, adds voiceovers, and synchronizes audio with subtitles. This multi-agent system enables fully automated, high-quality animated storytelling.}
    \label{fig:model-architecture}
    \vspace{-0.5em}
\end{figure*}

\vspace{-0.5em}
\section{Related Work}

\paragraph{Storyboard Visualization}
Storyboard visualization bridges script to video by generating coherent image sequences, demanding flexible, reusable character-background modules for multiple scenes. While adapter-based techniques like IP-Adapter \cite{ye2023ip}, Mix-of-Show \cite{gu2023mix}, T2I-Adapter \cite{mou2024t2i}, ROICtrl \cite{gu2024roictrl}, and StoryAdapter \cite{mao2024story} realize character consistency, they often overlook background continuity. Consistency-aware methods try to address this challenge. StoryGen \cite{liu2024intelligent} iteratively generates images using previous visual-language context, while StoryDiffusion \cite{zhou2024storydiffusion} employs a training-free Consistent Self-Attention module to improve feature alignment. Despite these advancements, the precise preservation of visual details, especially with multiple characters or changing backgrounds, remains a challenge.

\vspace{-0.5em}
\paragraph{Video Generation} 
Driven by the success of diffusion models, VDM \cite{ho2022video} pioneers their application to video generation, leading to improved models like Stable Video Diffusion \cite{blattmann2023stable}, ModelScope \cite{wang2023modelscope}, VideoCrafter1 \cite{chen2023videocrafter1}, and VideoCrafter2 \cite{chen2024videocrafter2}. While these models excel at animating single text or image, they struggle with maintaining consistency across multiple video clips. The emergence of Sora \cite{liu2024sora} has spurred a demand for coherent, consistent, and controllable storytelling videos. In response, several works\cite{wu2024mind, kim2024fifo, dalal2025one, guo2025long} have emerged to enhance storytelling video generation by improving temporal modeling and contextual understanding. Despite these advancements, challenges persist as single-pass long video generation often suffers from inconsistencies and repetitive content over time.  Considering this, models like HunyuanVideo \cite{kong2024hunyuanvideo}, Wan \cite{wan2025wan}, and CogVideoX \cite{yang2024cogvideox} still prioritize the effective utilization of text prompts and reference images, resulting in more faithful generation and improved motion depiction. Industry solutions \cite{bao2024vidu, PikaLabsWebsite, PixverseAIWebsite, RunwayAIWebsite} extend these advancements, enabling features like start/end frame-specific control.

\vspace{-0.5em}
\paragraph{Storytelling Video Agent}
Storytelling video generation typically follows a pipeline: script → keyframes → video clips → final video composition. Large multimodal models (LMMs) often act as high-level planners to coordinate each stage (e.g., \cite{GoogleGeminiFlash2.0, hurst2024gpt, li2025uni}). Frameworks such as VideoDirectorGPT \cite{lin2023videodirectorgpt}, DreamStory \cite{he2024dreamstory}, AnimDirector \cite{li2024anim}, MM-StoryAgent \cite{xu2025mm}, MovieAgent \cite{wu2025automated}, and DreamFactory \cite{xie2024dreamfactory} adhere to this conventional pipeline. Notably, certain tasks like scriptwriting, character design, voiceovers, and sound effects are typically handled manually during pre- or post-processing.

\vspace{-0.5em}
\section{Methodology} 

Our proposed framework, AniMaker, automates the creation of foundational storytelling animation from text. It uses a multi-agent system that mirrors professional animation pipelines, incorporating novel components for efficient candidate generation and comprehensive evaluation. The overall architecture is detailed in Figure~\ref{fig:model-architecture}.

\vspace{-0.5em}
\subsection{Task Formulation}

Automated generation of storytelling animation $\mathbf{V}_{\text{final}}$, from text input $\mathbf{T}_{\text{prompt}}$, can be represented as:
$$ \mathcal{F} : \mathbf{T}_{\text{prompt}} \rightarrow \mathbf{V}_{\text{final}} $$
This transformation is realized by mapping through several crucial intermediate representations:

\begin{itemize}[leftmargin=*]
    
    \item \textbf{Script ($\boldsymbol{\mathcal{P}}_{\text{script}}$):} Derived from $\mathbf{T}_{\text{prompt}}$, the script 
    $$\boldsymbol{\mathcal{P}}_{\text{script}} = ((\mathbf{shot}_k)_{k=1}^{\mathbf{N}_{\text{clips}}}, \mathbf{K}_{cut})$$
    defines:

    \begin{itemize}

        \item An ordered sequence of $N_{\text{clips}}$ shots, where each $\mathbf{shot}_k = (\mathbf{d}_k, \mathbf{C}_k, \mathbf{B}_k)$ specifies its textual description $\mathbf{d}_k$, involved characters $\mathbf{C}_k$, and background $\mathbf{B}_k$.

        \item A set of indices $\mathbf{K}_{cut} \subseteq \{1, \dots, \mathbf{N}_{\text{clips}}\}$, marking shot transitions, indicating where KeyFrame guidance is required for the new shot.

    \end{itemize}

    \item \textbf{Storyboard($\boldsymbol{\mathcal{S}}_{\text{board}}$):} Storyboard is established from $\boldsymbol{\mathcal{P}}_{\text{script}}$, containing:

    \begin{itemize}
    
        \item Character Bank: $\boldsymbol{\mathcal{B}}_{\text{char}} = \{ (\boldsymbol{c}, \mathbf{I}_c^{\text{char}}) \mid \boldsymbol{c} \in \boldsymbol{\mathcal{C}}_{\text{total}} \}$, with $\mathbf{I}_c^{\text{char}}$ as the reference image for character $\boldsymbol{c}$, where $\boldsymbol{\mathcal{C}}_{\text{total}}$ represents all available characters in the story.
        
        \item Background Bank: $\boldsymbol{\mathcal{B}}_{\text{bg}} = \{ (\boldsymbol{b}, \mathbf{I}_{b}^{\text{bg}}) \mid \boldsymbol{b} \in \boldsymbol{\mathcal{B}}_{\text{total}} \}$, with $\mathbf{I}_b^{\text{bg}}$ as the reference image for background $\boldsymbol{b}$, where $\boldsymbol{\mathcal{B}}_{\text{total}}$ represents all available backgrounds in the story.
        
        \item KeyFrames: for each $\mathbf{shot}_k$ where $k \in \mathbf{K}_{cut}$, a KeyFrame $\mathbf{F}_k^{\text{key}}$ is defined:
        $$ \mathbf{F}_k^{\text{key}} = \text{Visualize}(\{\mathbf{I}_c^{\text{char}} \mid \boldsymbol{c} \in \mathbf{C}_k\}, \mathbf{I}_{\mathbf{B}_k}^{\text{bg}}, \mathbf{d}_k) $$
        where $\text{Visualize}(\cdot)$ generates a keyframe by integrating the specified characters, background, and text description into a multimodal prompt.

    \end{itemize}

    \item \textbf{Video Clip Sequence ($\mathbf{v}$):} $\mathbf{v} = (\mathbf{v}_1, \dots, \mathbf{v}_{N_{\text{clips}}})$ is generated, corresponding to $(\mathbf{shot}_k)_{k=1}^{N_{\text{clips}}}$. Let $G_K$ and $G_C$ be abstract generative processes of each clip $\mathbf{v}_k$:

    \begin{itemize}

        \item If $k \in \mathbf{K}_{cut}$ (after shot transition): $\mathbf{v}_k \sim G_K(\mathbf{F}_k^{\text{key}}, \mathbf{d}_k)$.

        \item If $k \notin \mathbf{K}_{cut}$: $\mathbf{v}_k \sim G_C(\text{last\_frame}(\mathbf{v}_{k-1}), \mathbf{d}_k)$.

    \end{itemize}

    \item \textbf{Final Video ($\mathbf{V}_{\text{final}}$):} The assembled clip sequence $\text{Assemble}(\mathbf{v})$ then undergoes post-processing, utilizing information from $\boldsymbol{\mathcal{P}}_{\text{script}}$ to produce the polished $\mathbf{V}_{\text{final}}$. This may include additions like voiceovers and subtitles.
    $$ \mathbf{V}_{\text{final}} = \text{PostProcess}(\text{Assemble}(\mathbf{v}), \boldsymbol{\mathcal{P}}_{\text{script}}) $$

\end{itemize}

\vspace{-0.5em}
\subsection{Pipeline Overview}

AniMaker transforms textual input ($\mathbf{T}_{\text{prompt}}$) into compelling storytelling animation ($\mathbf{V}_{\text{final}}$) through four specialized agents working collaboratively. Central to this pipeline is the use of LLM prompt engineering by its four specialized agents, and we refer to the appendix for related details. The Director Agent (Section~\ref{sec:director-agent}) creates a detailed storyboard ($\boldsymbol{\mathcal{S}}_{\text{board}}$), the Photography Agent (Section~\ref{sec:photography-agent}) generates candidate video clips ($\mathbf{v}_k$) using MCTS-Gen (efficiently exploring generation space conserving computational resources), the Reviewer Agent (Section~\ref{sec:reviewer-agent}) evaluates clips with AniEval (our context-aware evaluation framework), and the Post-production Agent (Section~\ref{sec:post-production-agent}) assembles clips with voiceovers and subtitles.


\subsection{Multi-Agent Framework: AniMaker}
\subsubsection{\textbf{Director Agent}} 
\label{sec:director-agent} 
The Director Agent orchestrates storyboard generation through a two-stage process. First, Gemini 2.0 Flash \cite{GoogleGeminiFlash2.0} creates a raw script with shot descriptions ($\mathbf{shot}_k$), followed by automated validation for consistency and narrative flow. Second, in the Storyboard ($\boldsymbol{\mathcal{S}}_{\text{board}}$) realization phase, a Visual Bank is built: Character Bank is built ($\boldsymbol{\mathcal{B}}_{\text{char}}$) with Hunyuan3D \cite{zhao2025hunyuan3d} and Background Bank ($\boldsymbol{\mathcal{B}}_{\text{bg}}$) is built with FLUX1-dev \cite{FluxAIWebsite}. Then, GPT-4o \cite{hurst2024gpt} generates keyframes ($\mathbf{F}_k^{\text{key}}$) combining validated shot descriptions ($\mathbf{d}_k$) with Visual Bank imagery. This ensures visual consistency, with the resulting $\boldsymbol{\mathcal{S}}_{\text{board}}$ serving as the animation production blueprint in the following phases.

\vspace{-0.5em}
\subsubsection{\textbf{Photography Agent}} 
\label{sec:photography-agent} 

Converting storyboards to clips in multi-shot AI video generation presents challenges including distorted appearances, inconsistent motion, and object inconsistencies. 
Drawing from the "No Good" (NG) process in filmmaking, where numerous takes are recorded to achieve the perfect shot, we recognize the need to produce multiple candidate clips to identify the optimal one.
For optimal selection, each clip must not only possess high individual quality but also ensure consistency and coherence with both preceding and succeeding clips. However, naively generating $k$ candidates per clip creates a combinatorial explosion (e.g., $k^2$ for two-clip sequences). Notably, poor-quality current clips allow for pruning the search space, as further exploration down such paths is unlikely to be satisfactory.

\begin{figure*}[t]
    \vspace{-0.5em}
    \centering
    \includegraphics[width=0.9\textwidth]{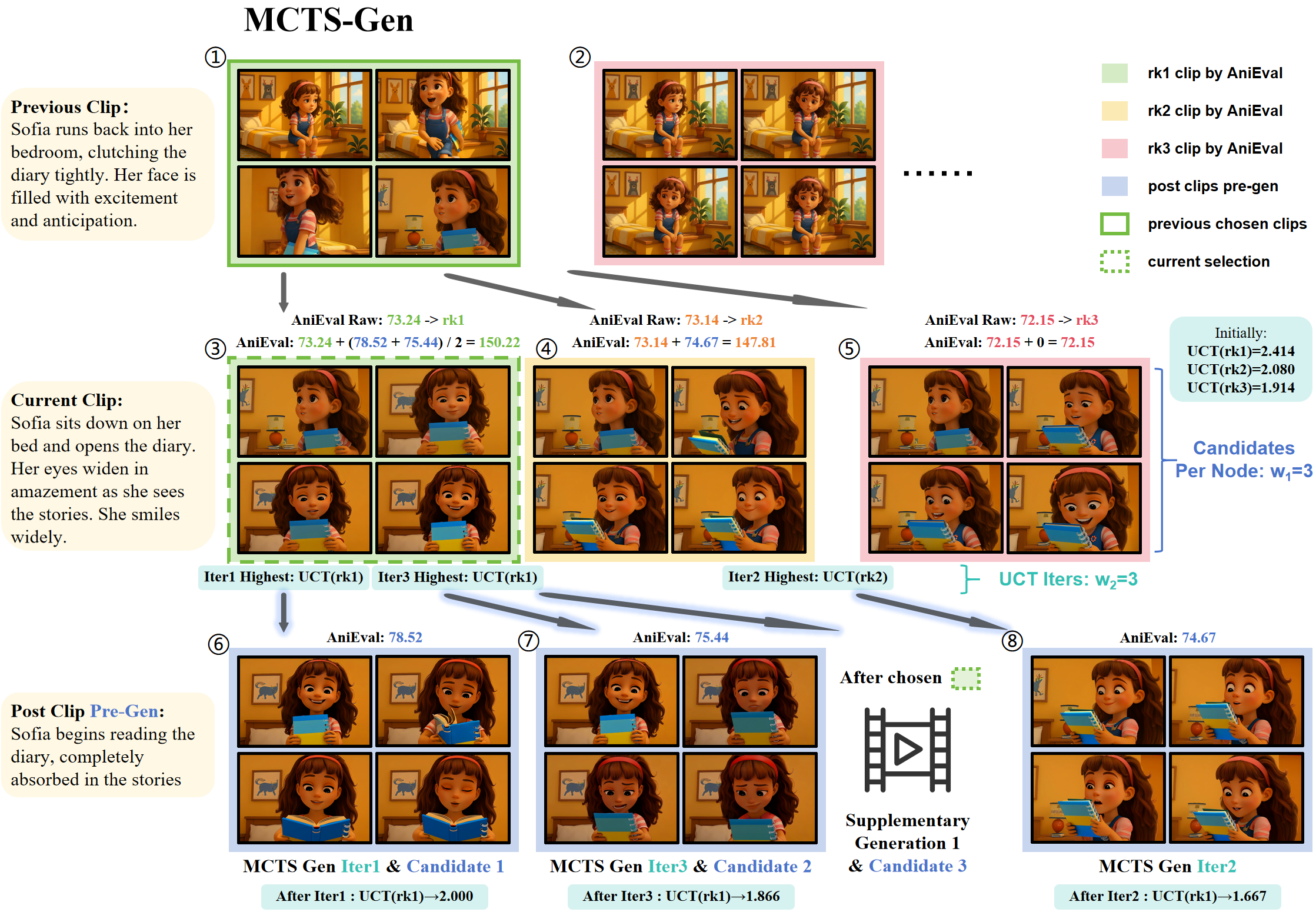}
    \caption{Illustration of our MCTS-Gen strategy for efficient Best-of-N Sampling.}
    \label{fig:mcts-architecture}
    \vspace{-0.5em}
\end{figure*}

Thus we propose \textbf{MCTS-Gen}, a Monte Carlo Tree Search (MCTS)-inspired method for multi-clip video generation (Figure~\ref{fig:mcts-architecture}). MCTS naturally fits this task: multi-clip sequences correspond to tree paths, with each clip as a node. Crucially, clip evaluation considers both intrinsic quality and inter-clip consistency, aligning with MCTS's backpropagation where child's scores update the parent's evaluation results. MCTS-Gen iteratively constructs a chosen path of selected video clips, extending this path by one clip per iteration. This process is controlled by the following parameters: $\mathbf{w_1}$ (initial candidate count per node), $\mathbf{w_2}$ (UCT-guided expansion times per iteration), and $\boldsymbol{\alpha}$ (exploration-exploitation balancing factor, default 1). Specifically, we use Wan 2.1 \cite{wan2025wan} for video clip generation, and the algorithm proceeds as follows:


1. \textbf{Expansion:} $\mathbf{w_1}$ (3 in Figure~\ref{fig:mcts-architecture}) initial child clips are generated from the chosen path's terminal node (node 1 in Figure~\ref{fig:mcts-architecture}). These clips (node 3, 4, 5 in Figure~\ref{fig:mcts-architecture}) are then scored using AniEval (detailed in Section~\ref{sec:reviewer-agent}) and ranked.

2. \textbf{Simulation:} Further $\mathbf{w_2}$ (3 in Figure~\ref{fig:mcts-architecture}) expansions are generated guided by the \textbf{UCT score}:
$$ \boldsymbol{UCT}(\boldsymbol{node_j}) = \frac{2.0}{\boldsymbol{rank}(\boldsymbol{node_j}) + 1} + \boldsymbol{\alpha} \cdot \sqrt{\frac{2.0}{\boldsymbol{child\_count}(\boldsymbol{node_j}) + 1}} $$
where $\boldsymbol{rank}$ is from the initial AniEval scoring, $\boldsymbol{child\_count}$ is dynamically updated, and $\boldsymbol{\alpha}$ balances exploitation and exploration. The node with the highest \textbf{UCT score} generates a new child clip.


3. \textbf{Backpropagation:} AniEval scores of child clips (nodes 6, 7, 8 in Figure~\ref{fig:mcts-architecture}) propagate upwards. A parent node's score is updated by adding the average score of children to its own.

4. \textbf{Selection:} The node with the highest AniEval score (node 3 in Figure~\ref{fig:mcts-architecture}) is added to the chosen path, then generating new child clips until reaching a total of $\mathbf{w_1}$ children, allowing the iterative generation process to continue.

This MCTS-Gen process systematically balances the exploration of diverse video generation space with the exploitation of promising paths, aiming to efficiently construct a high-quality, coherent video sequence while managing computational resources.

\vspace{-0.5em}
\subsubsection{\textbf{Reviewer Agent}}
\label{sec:reviewer-agent}

Existing objective metrics, such as CLIP Score and Inception Score, may identify superior video generation models, but often struggle to differentiate among candidates generated by the same model. Similarly, the widely adopted VBench has significant limitations. For instance, some of its metrics, like "dynamic degree", are overly simplistic, merely measuring pixel changes rather than accurately reflecting character action. More critically, VBench's "Consistency" related metrics, based on single-clip segmentation, prove unsuitable for multi-shot animation, which inherently involves frequent character and scene changes.

To address evaluation challenges, we introduce \textbf{AniEval}, a comprehensive evaluation framework built on EvalCrafter \cite{liu2024evalcrafter}. AniEval refines EvalCrafter's metrics for fully automated evaluation, \textit{e.g.}, by automating action assessment through comparing prompted character actions with those identified in video clips. 

Furthermore, responding to the specific demands of evaluating multi-shot animation characterized by multiple characters and diverse scenes, AniEval introduces several additional metrics: \textbf{DreamSim} \cite{fu2023dreamsim} assesses overall frame consistency; \textbf{Count-Score} \cite{cheng2024theatergen} aims at the issue of objects appearing or disappearing between shots; \textbf{Face Consistency} evaluates animated character facial consistency by training an \textit{InceptionNext} \cite{yu2024inceptionnext} model on the Anime Face Dataset \cite{AnimeFaceDatasetKaggle}, overcoming the limitations of conventional face recognition methods like MTCNN \cite{ku2020face} in anime face detection and tracking.

In conclusion, AniEval comprises 4 primary domains with 14 fine-grained metrics for comprehensive assessment (Table~\ref{tab:AniEval_metrics}). Additionally, AniEval supports contextual scoring by evaluating clips based on preceding and succeeding content, providing robust evaluation for multi-shot animation generation. Further implementation details of AniEval can be found in the appendix. 


\renewcommand{\arraystretch}{1.15}
\begin{table}[htbp]
\centering
\hspace{-0.5cm}
\caption{Introduction of AniEval Metrics.}
\label{tab:AniEval_metrics}
\footnotesize 
\setlength{\tabcolsep}{4pt} 
\begin{tabular}{@{}p{2.2cm}p{2.35cm}p{3.4cm}@{}}
\toprule
\textbf{Domain} & \textbf{Metric} & \textbf{Brief Description} \\
\midrule
Overall Video Quality & VQA\_A & Aesthetic video quality \\
& VQA\_T & Technical video quality \\
& MusIQ & Frame quality score \\
\midrule
Text-Video Alignment & Text-Video Consistency & Measured by CLIP \\
& Text-Story Consistency & Measured by BLIP-BLEU \\
& Detection-Score & Object generation accuracy \\
& Count-Score & Key object count accuracy \\
\midrule
Video Consistency & DreamSim & Perceptual frame-frame similarity \\
& Face Consistency & Animated character \newline facial consistency\\
& Warping Error & Temporal inconsistency \newline via pixel differences \\
& Semantic Consistency & Temporal semantic coherence \newline via CLIP \\
\midrule
Motion Quality & Action Recognition & Actions-prompt consistency \\
& Action Strength & Motion intensity via Flow-Score \\
& Motion AC-Score & Motion amplitude-prompt \newline consistency \\
\bottomrule
\end{tabular}
\end{table}
\renewcommand{\arraystretch}{1.0}

\vspace{-0.5em}
\subsubsection{\textbf{Post-production Agent}}
\label{sec:post-production-agent}
The Post-production Agent transforms video clip sequences into a polished animation film through three stages. Firstly, it leverages Gemini 2.0 Flash to generate a detailed voiceover script specifying narration, dialogue, emotional tones, and desired voice timbres. The agent then selects appropriate voice profiles based on character attributes (age, gender) and assesses text length for audio-visual synchronization. Secondly, the script is processed through CosyVoice2 \cite{du2024cosyvoice} to generate audio tracks, which undergo verification regarding duration consistency and content accuracy. Finally, the agent employs the MoviePy library for film assembly, integrating validated subtitles and performing comprehensive editing to ensure precise synchronization between visuals, voiceovers, and subtitles.


\vspace{-0.5em}
\section{Experiments} 

\subsection{Settings}

\subsubsection{Datasets}
To evaluate our AniMaker, we sample 10 narratives from TinyStories \cite{eldan2023tinystories}. These narratives feature complex multi-character interactions across diverse backgrounds, providing an ideal testbed for multi-shot animation generation.

\vspace{-0.5em}
\subsubsection{Baseline}
We evaluate several state-of-the-art storytelling models: StoryGen, StoryDiffusion, and StoryAdapter (visual narrative specialists), alongside MovieAgent, MMStoryAgent, and VideoGen-of-Thought (video generators). The latter group utilizes their built-in video modules, while StoryDiffusion and StoryAdapter are paired with external image-to-video models (CogVideoX and Wan 2.1).

\begin{table}[t]
\centering
\caption{Keyframe evaluation on Contextual Coherence
(Coherence), Image-Image Similarity (I-I Sim), and Text-Image Similarity
(T-I Sim).}
\label{tab:keyframe_evaluation}
\begin{tabular}{@{}lccc@{}}
\toprule
\textbf{Model} & \textbf{Coherence$\uparrow$} & \textbf{I-I Sim$\uparrow$} & \textbf{T-I Sim$\uparrow$} \\
\midrule
StoryGen & 0.54 & 0.77 & 0.22 \\
StoryDiffusion & 0.70 & 0.80 & 0.25 \\
StoryAdapter & 0.78 & {\textbf{0.83}} & 0.25 \\
MovieAgent & 0.59 & 0.65 & 0.23 \\
MMStoryAgent & 0.78 & {\textbf{0.83}} & 0.26 \\
VideoGen-of-Thought & 0.71 & 0.77 & 0.23 \\
\textbf{AniMaker(Ours)} & {\textbf{0.81}} & {\textbf{0.83}} & {\textbf{0.31}} \\
\bottomrule
\end{tabular}
\end{table}

\vspace{-0.5em}
\subsubsection{Evaluation Metrics}
For keyframe generation, we evaluate text-to-image alignment and cross-image consistency. Metrics include Text-to-Image CLIP (Coherence) \cite{clip}, Image-to-Image Similarity (I-I Sim) \cite{gal2022image}, and Text-Image Similarity (T-I Sim) \cite{hessel2021clipscore}. Video generation is assessed using VBench and our AniEval for comprehensive evaluation.

\vspace{-0.5em}
\subsection{Qualitative Analysis}
\begin{figure*}
    \centering
    \includegraphics[width=1.00\textwidth]{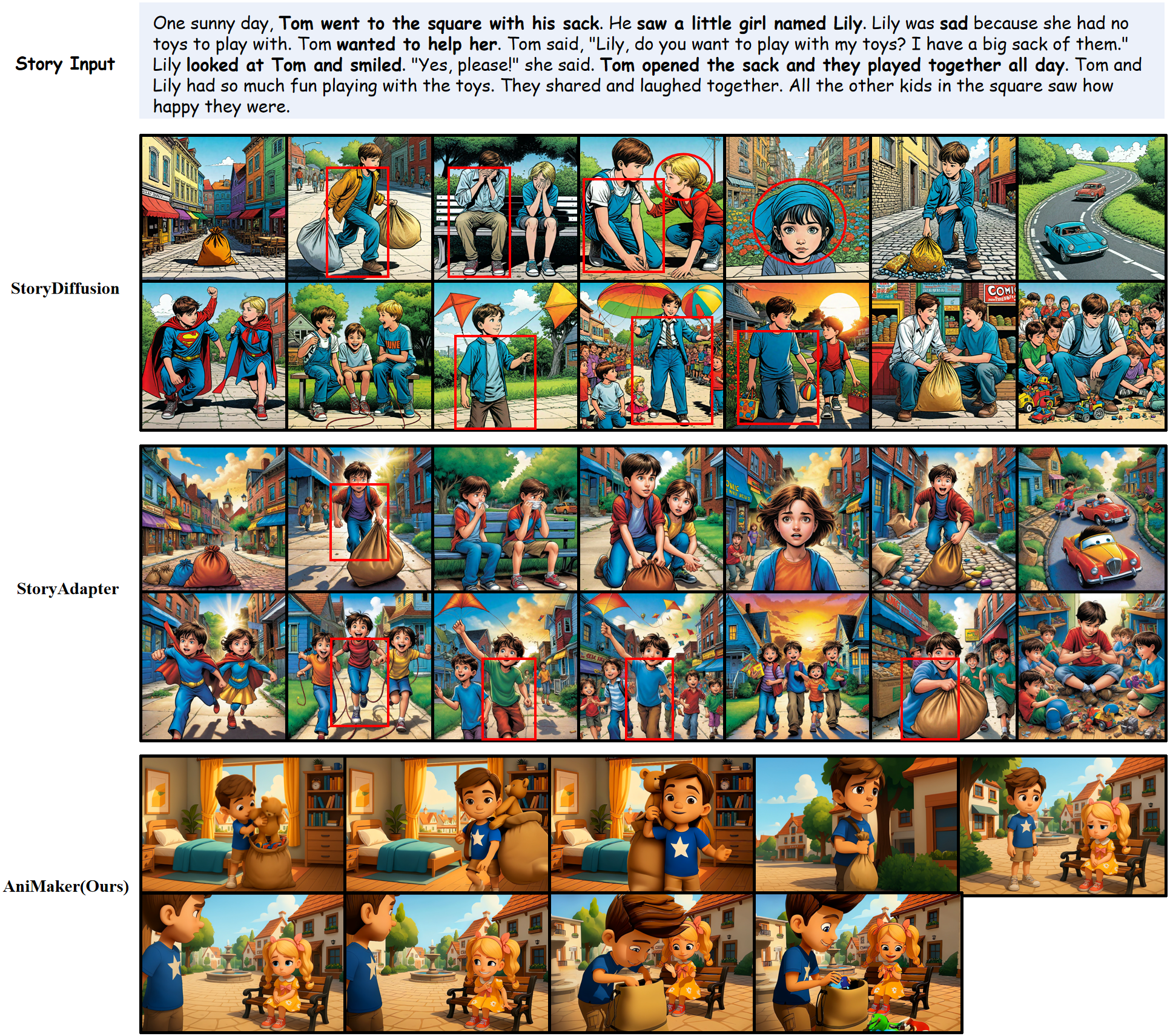}
    \caption{A comparative case showcasing AniMaker and models specialized in visual narratives. This figure illustrates the visualization of the short story of Tom and Lily. In the story, Tom brings a sack of toys to the town square, where he meets a sad girl named Lily who has no toys. Tom offers to share his toys, and the two children happily play together. Three models—StoryDiffusion, StoryAdapter, and AniMaker (ours)—are compared. AniMaker demonstrates superior narrative consistency, emotional expression, and character continuity across frames. It coherently depicts the extended action sequence of Tom picking up the sack, leaving his house, and arriving at the square. In contrast, while StoryDiffusion and StoryAdapter capture key moments from the story, they suffer from inconsistencies in visual coherence and character alignment, with mismatched character appearances highlighted by red boxes in the figure.}
    \label{fig:Tom&Lily}
\end{figure*}

\begin{figure*}
    \centering
    \includegraphics[width=1.00\textwidth]{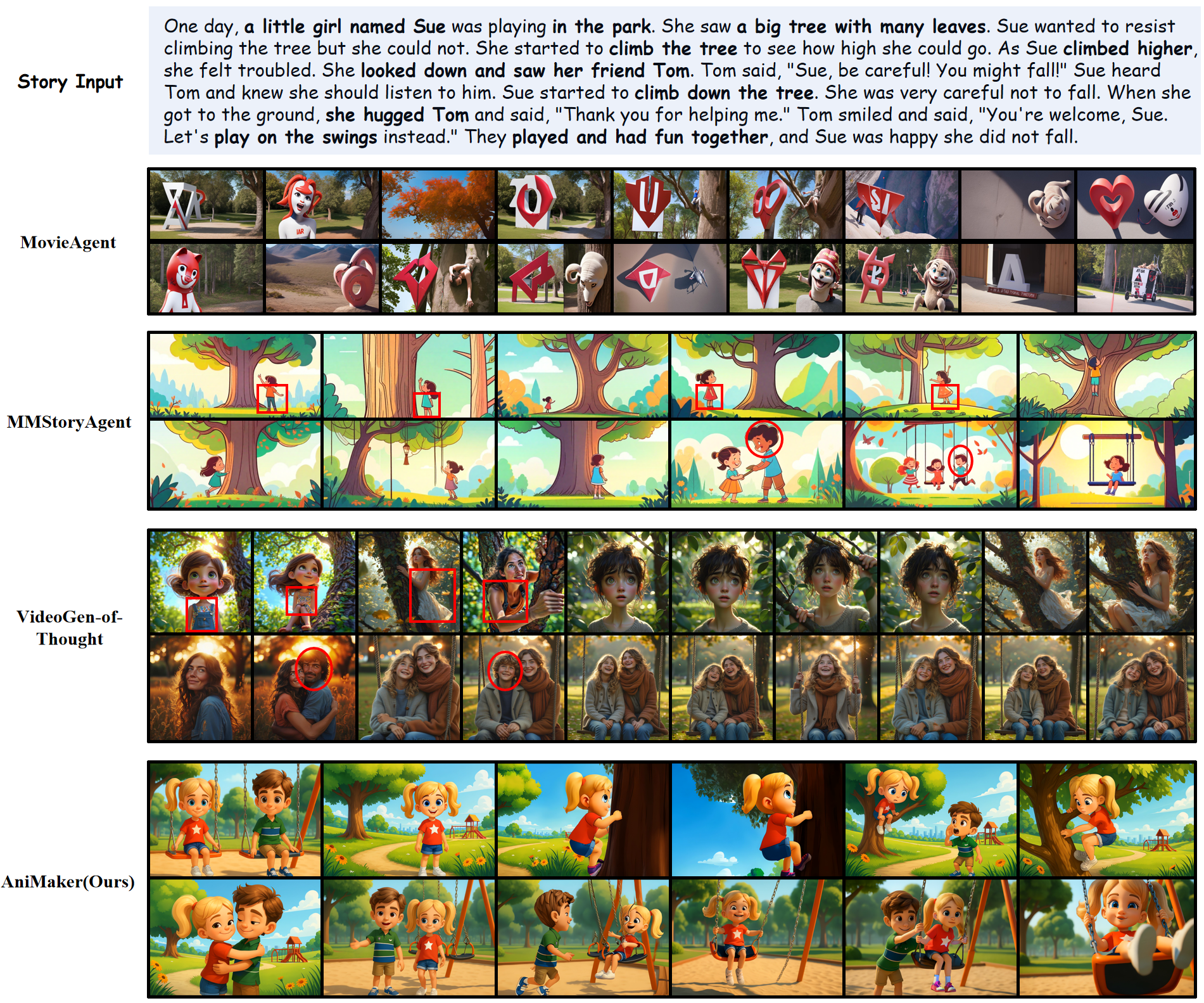}
    \caption{A comparative case showcasing AniMaker and models capable of generating storytelling videos. This figure visualizes the story of Sue, a little girl who tries to climb a big tree in the park but gets scared. Her friend Tom warns her to be careful, and she climbs down safely. Grateful, Sue hugs Tom, and they play on the swings together. The comparison includes MovieAgent, MMStoryAgent, VideoGen-of-Thought, and AniMaker (ours). AniMaker stands out with coherent scene progression, expressive character interactions, and consistent character identities. It clearly captures Sue’s emotional journey and key events—from climbing the tree and feeling afraid, to receiving help and having fun—demonstrating strong temporal and narrative alignment. In contrast, MovieAgent shows limited relevance to the input story, with inconsistent visuals and abstract content. VideoGen-of-Thought and MMStoryAgent follow the narrative more closely but still suffer from visual continuity issues, with character mismatches highlighted in red boxes.}
    \label{fig:Sue&Tom}
\end{figure*}

Based on the samples depicted in Figures~\ref{fig:Tom&Lily} and~\ref{fig:Sue&Tom}, we conduct a qualitative analysis of AniMaker's output, with a focus on visual fidelity and narrative coherence in multi-character, multi-scene animated storytelling.

\vspace{-0.5em}
\paragraph{Enhanced Consistency}
AniMaker demonstrates exceptional visual consistency across scenes. While baseline methods often face challenges in maintaining character and background consistency during scene transitions, AniMaker successfully preserves visual characteristics, even when switching back and forth between distinct scenes. This can be attributed to the Director Agent’s storyboard creation process, which provides a reliable set of reference images for both characters and backgrounds. Additionally, AniEval plays a crucial role in ensuring narrative coherence by incorporating a meticulous clip selection process, which evaluates consistency related metrics across clips.

\vspace{-0.5em}
\paragraph{Improved Action Representation}
AniMaker excels at depicting complex and extended character actions. Baseline methods often produce incomplete movements, particularly for multi-step action sequences (e.g., squatting, picking up an object, standing up, and walking away in Figure~\ref{fig:Tom&Lily}). Our MCTS-Gen strategy enables the Photography Agent to explore and select clip sequences that concatenate into coherent, complete long actions.

\vspace{-0.5em}
\paragraph{Seamless Transitions Between Clips}
AniMaker achieves smoother video transitions through an effective generation and selection mechanism. Central to this process is the Reviewer Agent, which leverages our AniEval framework to maximize visual continuity. By integrating cross-clip consistency metrics such as DreamSim and applying contextual scoring to adjacent clips, the agent effectively minimizes jarring visual disruptions between shots.

In summary, AniMaker outperforms existing methods through superior visual consistency across scenes, effective depiction of complex action sequences, seamless inter-clip transitions, and robust handling of multi-character, multi-scene narratives.

\begin{table}[t]
\centering
\caption{VBench evaluation results, presenting scores for Image Quality (I.Q.), Semantic Consistency (S.C.), Background Consistency (B.C.), Animation Quality (A.Q.), Motion Smoothness (M.S.), Dynamic Degree (D.D.), and Average Rank (Rk. Avg. - the average ranking position across all models).}
\label{tab:vbench_evaluation}
\small
\setlength{\tabcolsep}{1pt} 
\begin{tabular}{@{}lccccccc@{}}
\toprule
\textbf{Model} & \textbf{I.Q.$\uparrow$} & \textbf{S.C.$\uparrow$} & \textbf{B.C.$\uparrow$} & \textbf{A.Q.$\uparrow$} & \textbf{M.S.$\uparrow$} & \textbf{D.D.$\uparrow$} & \textbf{Rk. Avg.$\downarrow$} \\
\midrule
StableDiffusion+Cog. & 75.52 & 78.05 & 85.27 & 59.61 & 97.63 & 33.58 & 5.83 \\
StableDiffusion+Wan. & \underline{76.93} & 78.54 & 87.43 & 69.75 & 96.67 & 60.30 & 4.33 \\
StoryAdapter+Cog. & 76.17 & 72.17 & 88.03 & 63.55 & 98.16 & 26.71 & 5.33 \\
StoryAdapter+Wan. & 75.96 & 75.04 & 88.64 & 73.38 & 97.02 & \textbf{84.73} & 4.17 \\
MovieAgent & 72.09 & 68.61 & 79.84 & 55.40 & 99.01 & 35.44 & 6.33 \\
MMStoryAgent & 76.41 & \textbf{87.27} & \textbf{90.74} & \underline{73.84} & \textbf{99.80} & 0.00 & \underline{2.67} \\
VoT & 63.85 & 75.11 & 85.78 & \textbf{74.91} & \underline{99.25} & 3.50 & 4.83 \\
\textbf{AniMaker(Ours)} & \textbf{76.96} & \underline{84.27} & \underline{89.06} & 69.79 & 98.50 & \underline{66.97} & \textbf{2.50} \\
\bottomrule
\end{tabular}
\end{table}

\vspace{-0.5em}
\subsection{Quantitative Comparisons}
\begin{figure}[t]
    \centering
    \includegraphics[width=0.45\textwidth]{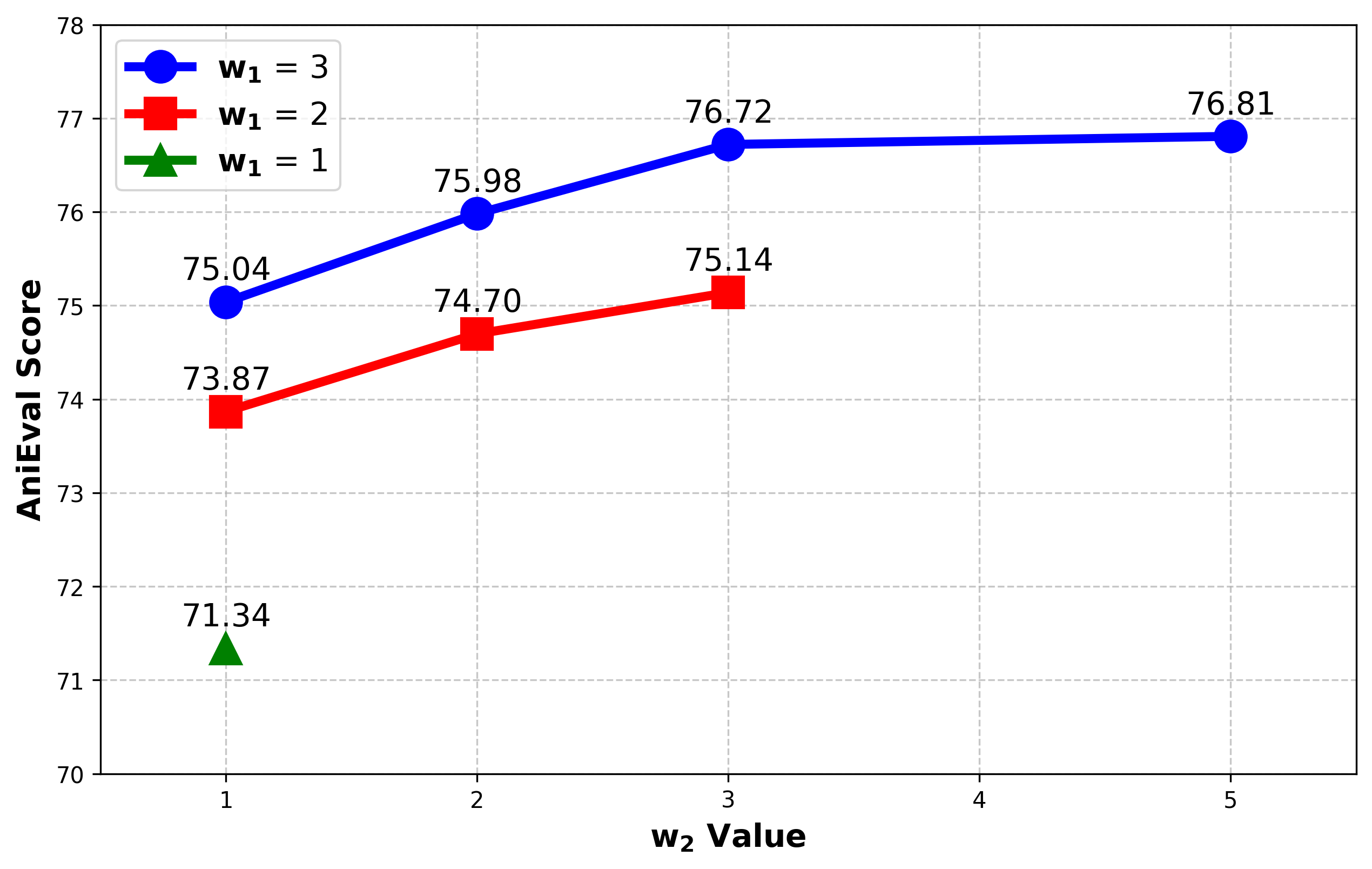}
    \caption{AniEval Score of Different $\mathbf{w_1}$ (initial candidate count) and $\mathbf{w_2}$ (expansion iterations) Combinations.}
    \label{fig:W1W2_Combinations}
\end{figure}

\paragraph{Keyframe Generation Analysis}
Keyframes are the core components of a storyboard. Table~\ref{tab:keyframe_evaluation} demonstrates that AniMaker outperforms all other competing methods across metrics. Notably, AniMaker achieves a Text-to-Image Similarity (T-I Sim) score of 0.31, representing a 19.2\% improvement over the best-performing baseline method. This advantage stems from our multimodal storyboard generation approach that incorporates character references, background references, and text prompts for keyframe generation, rather than relying solely on text.
\vspace{-0.5em}
\paragraph{VBench Evaluation Analysis}
Table~\ref{tab:vbench_evaluation} shows that AniMaker achieves the best average rank (2.50), demonstrating consistent top-tier performance across metrics. While MMStoryAgent excels in Semantic Consistency (S.C.) and Background Consistency (B.C.), its 0.00 score in Dynamic Degree (D.D.) reveals a critical limitation—it produces static image sequences resembling comic strips rather than true animation. This exposes VBench's limitations: it favors static consistency (where even brief static images can achieve high scores) and focuses on individual clips rather than multi-scene, multi-character animation. These findings highlight the need for a new evaluation framework specifically designed for storytelling animation quality.
\vspace{-0.5em}
\paragraph{AniEval Evaluation Analysis}
We introduce AniEval for better evaluation of storytelling Animation. Table~\ref{tab:anieval_evaluation} presents results from our AniEval framework. AniMaker outperforms all competitors with a total score of 76.72, representing a 14.6\% improvement over the second-best method (i.e. VideoGen-of-Thought with 66.93). Particularly noteworthy is AniMaker's exceptional performance in Video Consistency (V.C.), surpassing the best-performing baseline method by 15.5\%. While, AniMaker's relatively lower performance in Text-Video Alignment (T.V.A.) can be attributed to our agent framework's creative adaptation of stories into scripts, where additional narrative elements are introduced. Compared to VBench, AniEval's assessment results align remarkably better with human evaluations (Table~\ref{tab:model_evaluation}), demonstrating that AniEval provides a more comprehensive and accurate assessment of multi-shot animation quality than previous metrics.
\vspace{-0.5em}
\paragraph{MCTS-Gen Parameter Analysis}
Figure~\ref{fig:W1W2_Combinations} shows how MCTS-Gen parameters $\mathbf{w_1}$ (initial candidate count) and $\mathbf{w_2}$ (expansion iterations) affect generation quality. Two patterns emerge: higher $\mathbf{w_1}$ values improve AniEval scores by providing more initial candidates, while higher $\mathbf{w_2}$ values enhance performance by better evaluating each clip's suitability for continuous generation. Importantly, once certain thresholds are reached, configurations with fewer total generations (e.g., $\mathbf{w_1}=3, \mathbf{w_2}=3$ with 4.37 generations per node) perform comparably to those requiring more (e.g., $\mathbf{w_1}=3, \mathbf{w_2}=5$ with 5.76 generations per node). This demonstrates MCTS-Gen's efficiency—compressing the search space by over 50\% compared to exhaustive search strategies (which require 9 generations per node for the same candidates) while maintaining quality.

\begin{table}[htbp]
\centering
\caption{AniEval evaluation results, presenting scores for Overall Video Quality (O.V.Q.), Text-Video Alignment (T.V.A.), Video Consistency (V.C.), Motion Quality (M.Q.), and Total Performance.}
\label{tab:anieval_evaluation}
\small 
\setlength{\tabcolsep}{4pt} 
\begin{tabular}{@{}lccccc@{}}
\toprule
\textbf{Model} & \textbf{O.V.Q.$\uparrow$} & \textbf{T.V.A.$\uparrow$} & \textbf{V.C.$\uparrow$} & \textbf{M.Q.$\uparrow$} & \textbf{Total$\uparrow$} \\
\midrule
StoryDiffusion+CogVideoX & 46.54 & 86.05 & 47.14 & 70.35 & 56.75 \\
StoryDiffusion+Wan 2.1 & 47.07 & 84.99 & 47.13 & 71.00 & 56.55 \\
StoryAdapter+CogVideoX & 56.76 & \textbf{87.38} & 55.89 & 69.73 & 63.95 \\
StoryAdapter+Wan 2.1 & 60.39 & 86.99 & 51.41 & 72.11 & 62.37 \\
MovieAgent & 41.17 & 68.50 & 68.68 & 70.16 & 61.95 \\
MMStoryAgent & 47.93 & 75.27 & 63.54 & 61.39 & 62.79 \\
VideoGen-of-Thought & 66.17 & 72.95 & 65.42 & 66.72 & 66.93 \\
\textbf{AniMaker(Ours)} & \textbf{81.87} & 74.30 & \textbf{79.35} & \textbf{72.66} & \textbf{76.72} \\
\bottomrule
\end{tabular}
\end{table}

\vspace{-0.5em}
\subsection{Human Rating} 
Following MovieAgent's evaluation framework \cite{wu2025automated}, we conduct human evaluation with 10 participants on 90 storytelling videos from 9 models across 10 stories. Each video was rated on a 1-5 scale across five dimensions: Visual Appeal, Script Faithfulness, Narrative Coherence, Character Consistency, and Physical Law Adherence. Note that evaluating complete storytelling videos (rather than individual clips) typically yields lower scores due to the increased complexity and length. As shown in Table \ref{tab:model_evaluation}, our model achieves superior performance across all metrics, especially in terms of Character Consistency. 

\begin{table}[htbp]
\centering
\caption{Human rating results on a 1–5 scale, covering Character Consistency (C.C.), Narrative Coherence (N.C.), Physical-Law Adherence (P.L.), Script Faithfulness (S.F.), and Visual Appeal (V.A.).}
\label{tab:model_evaluation}
\small
\setlength{\tabcolsep}{3pt}
\begin{tabular}{@{}lcccccc@{}}
\toprule
\textbf{Model} & \textbf{C.C.$\uparrow$} & \textbf{N.C.$\uparrow$} & \textbf{P.L.$\uparrow$} & \textbf{S.F.$\uparrow$} & \textbf{V.A.$\uparrow$} & \textbf{Avg.$\uparrow$} \\
\midrule
StoryDiffusion+CogVideoX & 1.37 & 1.48 & 1.37 & 1.67 & 1.56 & 1.49 \\
StoryDiffusion+Wan 2.1 & 2.00 & 1.82 & 1.82 & 2.00 & 2.11 & 1.95 \\
StoryAdapter+CogVideoX & 1.64 & 1.39 & 1.46 & 1.68 & 1.71 & 1.58 \\
StoryAdapter+Wan 2.1 & 2.04 & 1.82 & 1.89 & 2.04 & 2.57 & 2.07 \\
MovieAgent & 1.19 & 1.26 & 1.44 & 1.26 & 1.48 & 1.33 \\
MM\_StoryAgent & 1.62 & 1.62 & 1.72 & 1.83 & 2.24 & 1.81 \\
VideoGenoT & 1.67 & 1.74 & 1.78 & 1.74 & 2.26 & 1.84 \\
\textbf{AniMaker(Ours)} & \textbf{3.44} & \textbf{3.24} & \textbf{3.04} & \textbf{3.08} & \textbf{3.28} & \textbf{3.22} \\
\bottomrule
\end{tabular}
\end{table}

\vspace{-0.5em}
\subsection{Ablation Studies}
We conduct two key ablation experiments. First, we ablate MCTS-Gen by setting $w_1=1, w_2=1$, essentially generating only one candidate per clip. This change results in a 7.1\% reduction in performance (the green triangle in Figure~\ref{fig:W1W2_Combinations}) on AniEval, confirming the importance of our MCTS-driven generation strategy. Notably, even this ablated version still outperforms the best-performing baseline method by 6.6\%, further demonstrating that our multi-agent framework remains highly competitive even without clip candidate selection. Next, we ablate AniEval by generating five candidates per clip with VBench for selection. This yields a score of 73.18, a 4.6\% decrease compared to our raw method. Qualitative assessment of the resulting videos also reveals noticeable degradation in action expressiveness and cross-clip consistency, highlighting the importance of AniEval for storytelling animation.

\vspace{-0.5em}
\section{Limitations}

Despite employing state-of-the-art models for storyboard and video clip generation, a significant gap remains between current model capabilities and the quality required for commercial film production. A primary limitation is the restricted narrative scope and complexity our method can handle. Compared to the intricate cinematic grammar and narrative structures utilized in professional filmmaking, as detailed in \cite{ronfard2021film}, our approach currently falls short.

These shortcomings are rooted in several underlying defects within the video generation models. They are prone to producing animation artifacts, along with issues of incoherence and hallucination that disrupt narrative continuity. Furthermore, the models exhibit a relatively poor adherence to fundamental physical laws, resulting in unrealistic interactions between characters and scene elements, as also highlighted in \cite{shi2025jailbreak}. Luckily, our framework is built on a modular, plug-and-play architecture that allows for seamless model integration. As more advanced generative models emerge, we are committed to continuously updating our framework and sharing these workflow improvements with the community to address these limitations.

\vspace{-0.5em}
\section{Conclusion}

We present AniMaker, a comprehensive multi-agent framework that transforms text input into coherent storytelling animation by emulating professional workflows. Our system introduces two key innovations: MCTS-Gen, which optimizes exploration-exploitation balance during clip generation, and AniEval, the first evaluation framework specifically designed for multi-shot storytelling animation. AniMaker orchestrates specialized agents that seamlessly collaborate across storyboarding, generation, evaluation, and post-production stages. Our quantitative results validate the effectiveness of this approach, with substantial gains in both technical metrics and perceived quality. These advances mark an important step toward bridging the gap between AI-generated content and professional animation standards, paving the way for more accessible and high-quality animated storytelling production.

\begin{acks}
We thank the editor and reviewers for their efforts in improving our paper. This work was supported by grants: Natural Science Foundation of China (No. 62422603), Guangdong Basic and Applied Basic Research and Foundation (No.2024B0101050003) and Shenzhen Science and Technology Progam (No. ZDSYS20230626091203008).
\end{acks}

\bibliographystyle{ACM-Reference-Format}
\bibliography{sample-bibliography}

\clearpage

\clearpage

\clearpage

\appendix

\section{ LMM Prompts }
\begin{figure*}
    \centering
    \includegraphics[width=1.00\textwidth]{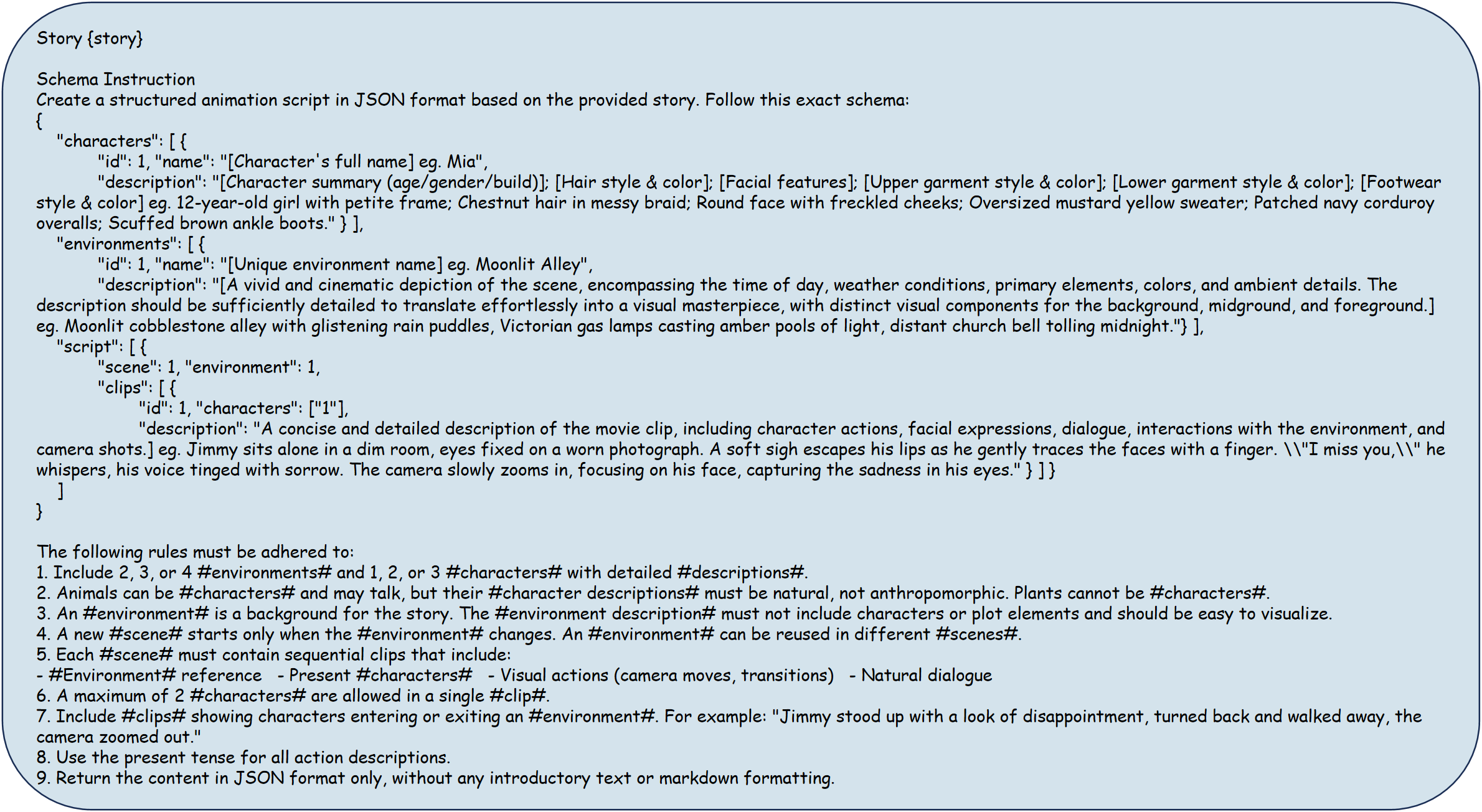}
    \caption{Prompt for Story to Script Generation}
    \label{fig:prompt1}
\end{figure*}

\begin{figure*}
    \centering
    \includegraphics[width=1.00\textwidth]{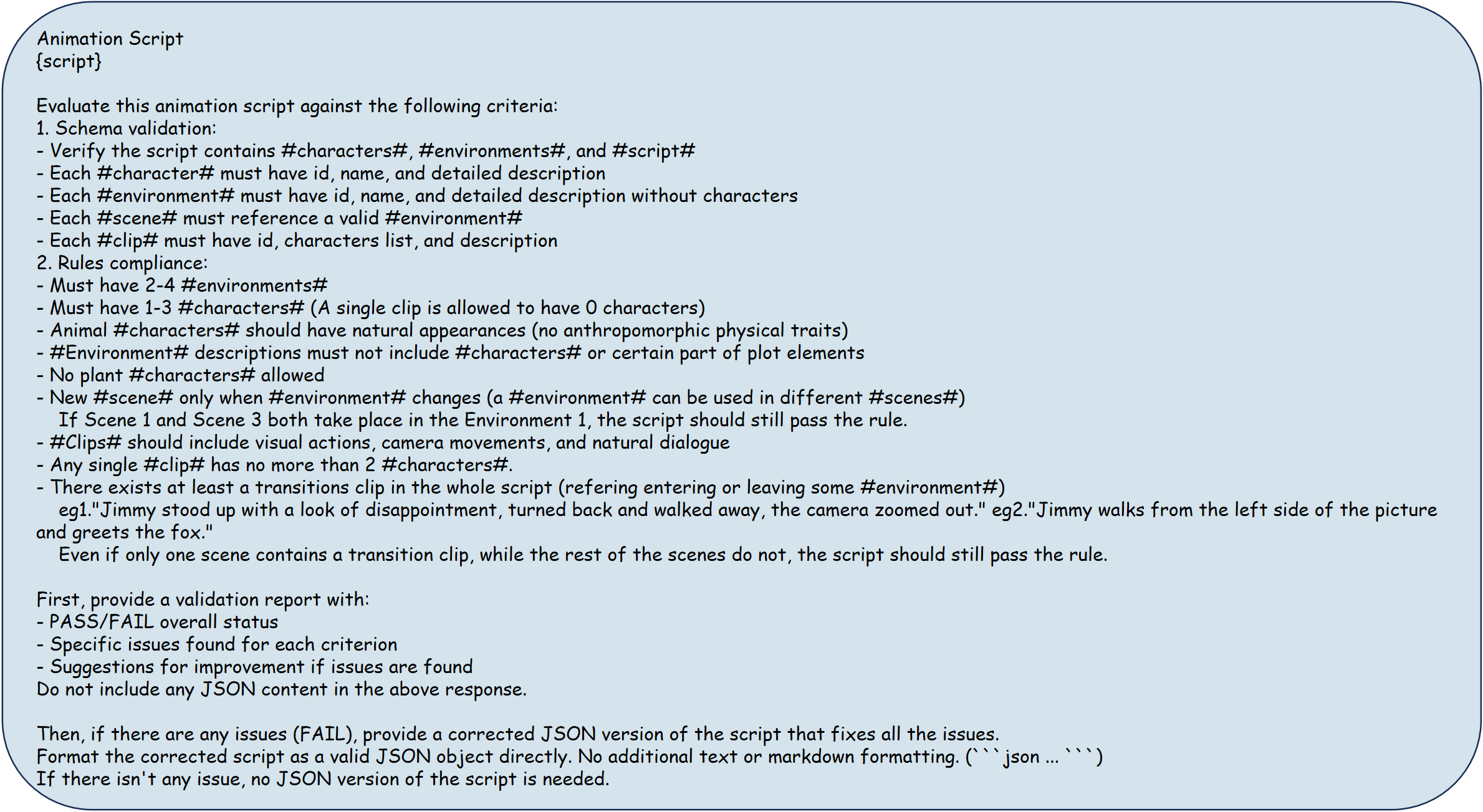}
    \caption{Prompt for Story to Script Check}
    \label{fig:prompt2}
\end{figure*}

\begin{figure*}
    \centering
    \includegraphics[width=1.00\textwidth]{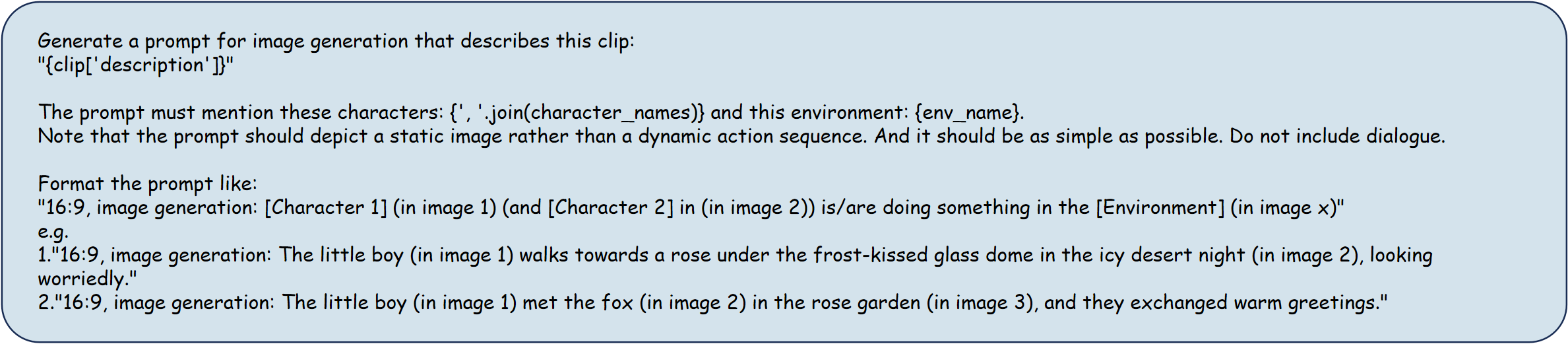}
    \caption{Prompt for Scene Image Generation}
    \label{fig:prompt3}
\end{figure*}

\begin{figure*}
    \centering
    \includegraphics[width=1.00\textwidth]{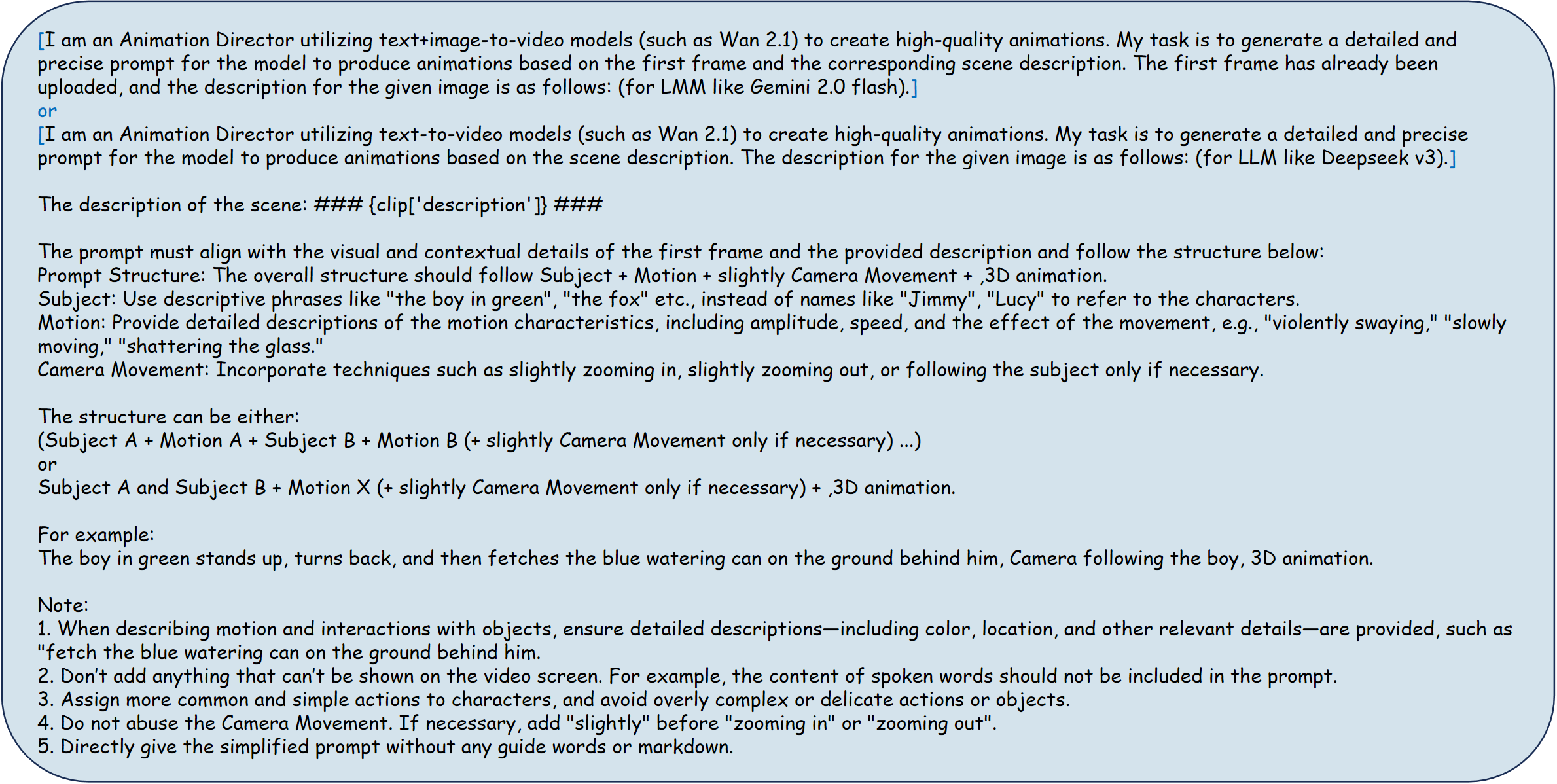}
    \caption{Prompt for Video Generation}
    \label{fig:prompt4}
\end{figure*}

\begin{figure*}
    \centering
    \includegraphics[width=1.00\textwidth]{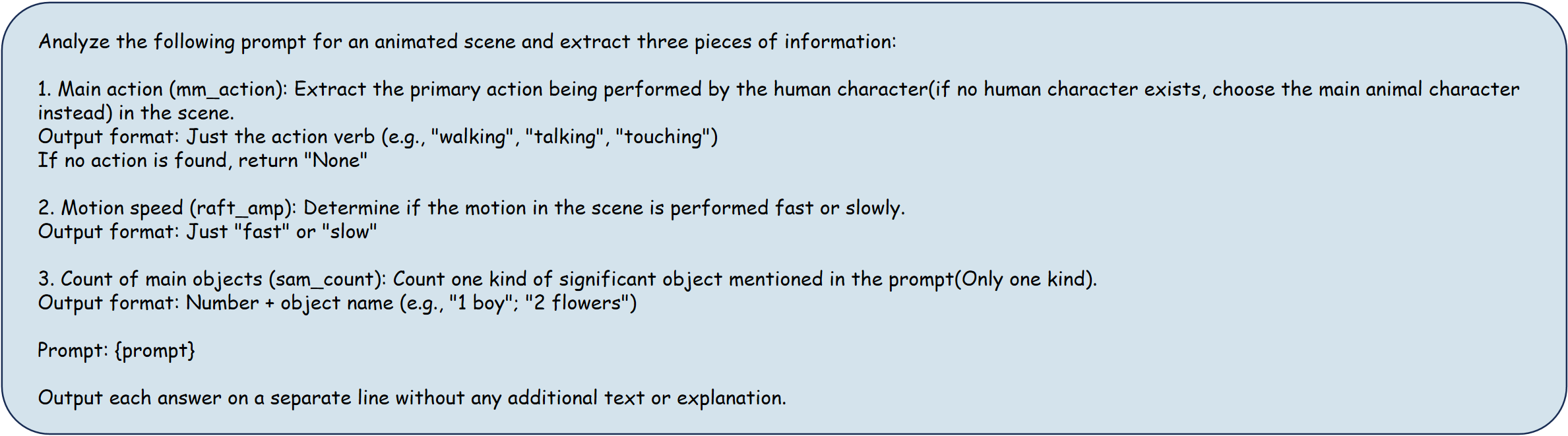}
    \caption{Prompt for Video Generation Prompt Analysis}
    \label{fig:prompt5}
\end{figure*}

\begin{figure*}
    \centering
    \includegraphics[width=1.00\textwidth]{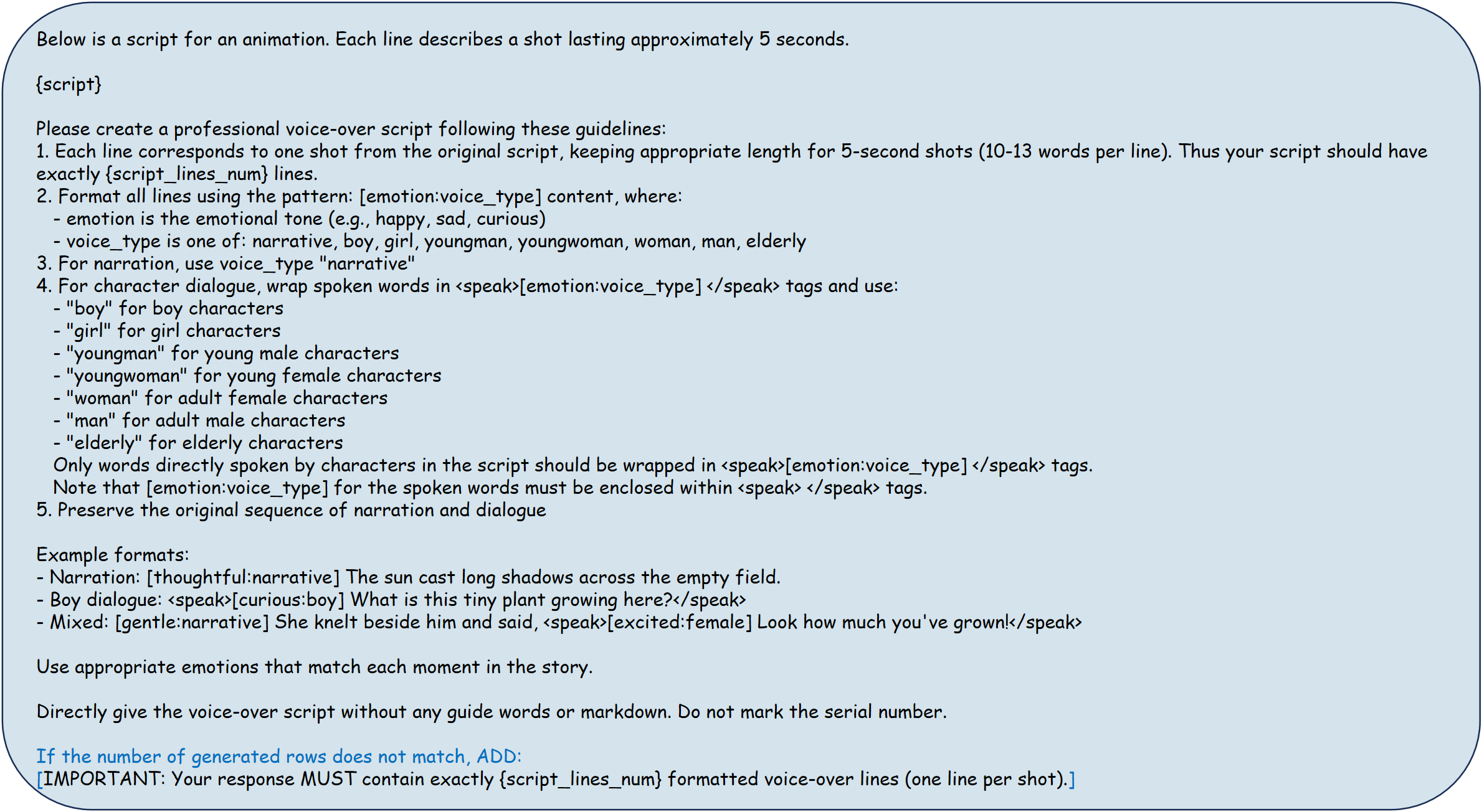}
    \caption{Prompt for Voiceover Script}
    \label{fig:prompt6}
\end{figure*}

\begin{figure*}
    \centering
    \includegraphics[width=1.00\textwidth]{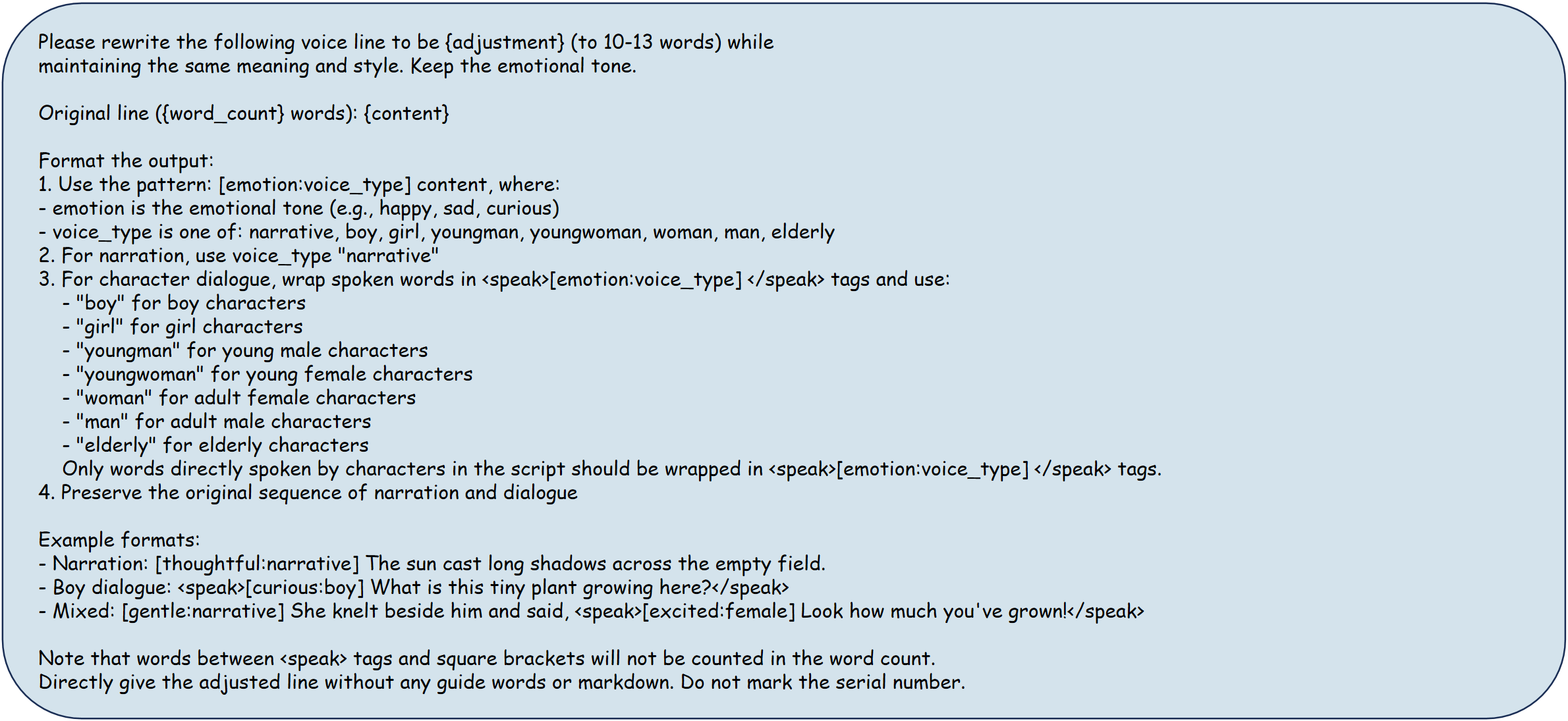}
    \caption{Prompt for Voiceover Script Modification}
    \label{fig:prompt7}
\end{figure*}

\begin{figure*}
    \centering
    \includegraphics[width=1.00\textwidth]{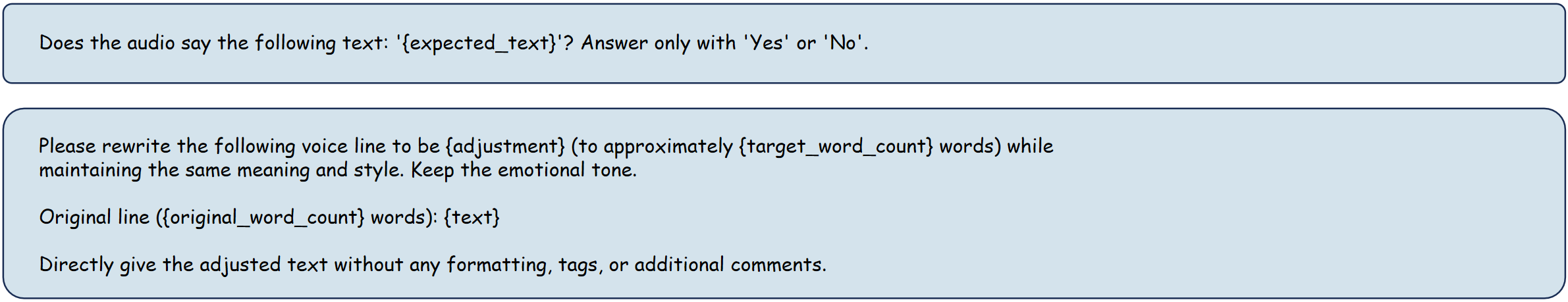}
    \caption{Prompt for Voiceover Script Check}
    \label{fig:prompt8}
\end{figure*}

This section presents the comprehensive set of prompts engineered for the LMM, which steers the entire automated animation creation pipeline. The figures below illustrate the sequence of these prompts, starting from the initial transformation of a narrative story into a structured script (Figure~\ref{fig:prompt1}) and its corresponding validation (Figure~\ref{fig:prompt2}). Following this, we detail the prompts for generating scene imagery (Figure~\ref{fig:prompt3}), the video itself (Figure~\ref{fig:prompt4}), and an analysis of the video generation prompt (Figure~\ref{fig:prompt5}). The final set of prompts covers the creation (Figure~\ref{fig:prompt6}), modification (Figure~\ref{fig:prompt7}), and verification (Figure~\ref{fig:prompt8}) of the voiceover script.

\section {Details of AniEval }
AniEval enhances the evaluation framework of EvalCrafter \cite{liu2024evalcrafter} through two main aspects. First, it automates the assessment of action related metrics by comparing user prompts with actions identified in the video clips. Second, it integrates several new metrics, including DreamSim \cite{fu2023dreamsim}, Count-Score \cite{cheng2024theatergen}, and \textbf{Face Consistency}. For a comprehensive assessment, AniEval is organized into 4 primary domains comprising 14 fine-grained metrics, as detailed below.

\subsection{Overall Video Quality}

\subsubsection{VQA\_A and VQA\_T}

Dover \cite{wu2023dover} is employed to assess generated video quality, providing both an aesthetic score (VQA\_A) and a technical score (VQA\_T). The aesthetic score focuses on the overall visual appeal of the video, assessing aspects such as composition, color harmony, and artistic quality, while technical score evaluates common distortions like noise and artifacts.

\subsubsection{MusIQ}

MUSIQ \cite{ke2021musiq} is adopted to evaluate the perceptual quality of the generated videos. Each clip is first sampled into a set of frames, after which the MUSIQ score is computed for every frame and averaged across the entire sequence. This yields a unified quality index that simultaneously captures global composition, local sharpness, and typical distortions such as blur and compression artifacts.

\subsection{Text-Video Alignment}

\subsubsection{Text-Video Consistency}

CLIPScore \cite{radford2021learning} is employed to measure the semantic consistency between the input text prompt and the generated video. The pretrained ViT-B/32 CLIP model serves as the feature extractor, yielding text embeddings and frame-wise image embeddings; their cosine similarity is computed for every frame. The overall CLIPScore is obtained by averaging these per-frame similarities across the entire sequence.

\subsubsection{Text-Story Consistency}

To further evaluate the alignment between the input prompt and the generated video, a method leveraging BLIP2 \cite{li2023blip} and BLEU \cite{papineni2002bleu} is also adopted. The BLIP2 model serves as a caption generator, producing five distinct text descriptions for each video. The BLEU score is then computed to measure the similarity between the original input prompt and each of these five generated captions. The overall score is the average of the five resulting BLEU scores.

\subsubsection{Detection-Score}

For each prompt, Gemini first identifies the key object and the desired quantity. To quantify its presence in the video, A Detection Score is computed by the SAMTrack model [cheng2023segment]. SAMTrack performs object detection on uniformly sampled frames; each frame returns a binary result (1 if the object is detected, 0 otherwise). The Detection Score is the mean of these binary values across all sampled frames.

\subsubsection{Count-Score}

Furthermore, to evaluate whether the video contains the correct number of objects, a Count Score is calculated. Using the same method, the number of detected objects in each frame is counted. This number is compared to the target quantity, yielding a binary frame-level score: 1 if the counts match, and 0 otherwise. The Count Score is the mean of these binary values across all sampled frames.

\subsection{Video Consistency}

\subsubsection{DreamSim}

DreamSim \cite{fu2023dreamsim} is employed to measure the overall similarity between a given clip and its adjacent clips. For each comparison (i.e., between the current clip and its preceding or succeeding clip), frames are extracted from both. For every frame in the current clip, its minimum distance to any frame in the adjacent clip is calculated by Dreamsim. The final DreamSim score for the pair is the average of these distances.

\subsubsection{Face Consistency}

The Face Consistency metric evaluates animated character facial consistency using an InceptionNext \cite{yu2024inceptionnext} model trained on the Anime Face Dataset \cite{AnimeFaceDatasetKaggle}. The process begins by employing a Segment-and-Track-Anything (SAM) based approach to detect and extract character faces from video clips and reference image, guided by a text prompt. Each extracted face is then passed through the InceptionNext model to obtain a high-dimensional feature embedding, where the L2 distance between embeddings serves as a measure of dissimilarity. The metric performs two types of evaluation: for Reference-to-Video Consistency, it calculates the average L2 distance between the reference image's face embedding and all face embeddings from the video frames. For Temporal Consistency between adjacent video clips featuring the same character, it determines the L2 distance from each face in the current clip to any face in the adjacent clip, and the final score is the average of these distances.

\subsubsection{Warping Error}

A pre-trained optical flow estimation network \cite{teed2020raft} is employed to obtain the optical flow between pairs of frames. Using this information, one frame is warped to align with the next. The pixel-wise difference between the warped image and the actual target frame is then calculated. Warp differences are computed for every frame pair, and the final score is obtained by averaging these results across the entire sequence.

\subsubsection{Semantic Consistency}

Beyond pixel-wise error, semantic consistency is also evaluated to assess the consistency between consecutive frames. Specifically, this involves calculating the cosine similarity between the feature embeddings (extracted by ViT-B/32 CLIP model) of each pair of adjacent frames. The final semantic consistency score is then derived by averaging these similarity values across all consecutive frame pairs in the video.

\subsection{Motion Quality}

For every prompt, Gemini is used to determine:

mm\_action: Describe the primary action of the main character – e.g., “walking”, “talking”, “hugging”.

raft\_amp: The perceived motion speed – “fast” or “slow”.

\subsubsection{Action Recognition}

The evaluation process begins by using the MMAction2 \cite{contributors2020openmmlab} toolbox and the VideoMAE V2 model \cite{wang2023videomae} to identify the top five most probable actions and their corresponding confidence scores. Subsequently, a CLIP model calculates the textual similarity between these five predictions and the target action (mm\_action). The final score is a weighted sum of these similarities, with the confidence scores serving as the weights.

\subsubsection{Action Strength}

The general action strength of the generated video is considered to ensure it is not overly static. To achieve this, the RAFT model \cite{teed2020raft} is first employed to extract the dense optical flows between each pair of frames. An average flow score is then computed by averaging these dense flows across the entire video clip.

\subsubsection{Motion AC-Score}

The alignment between a video's motion and the prompt is evaluated using the average optical-flow score. A classification of "fast" is assigned if the score exceeds the empirical threshold of $\rho$ = 5; otherwise, it is labeled "slow." A final comparison is then made between this label and the motion level required by the prompt.


\end{document}